\documentclass[a4]{article}
\usepackage{url}
\usepackage[pdftex]{graphicx}
\usepackage[english]{babel}

\begin{document}

\title{\textbf{OrfMapper: A Web-Based Application for Visualizing Gene Clusters on Metabolic Pathway Maps}\vspace{1cm}}

\author{Martin Vellguth \& R\"obbe W\"unschiers\footnote{to whom correspondence should be addressed}\\
Institute for Genetics, University of Cologne\\
Z\"{u}lpicher Strasse 47, D-50674 Cologne, Germany\vspace{2cm}}

\date{}

\maketitle

\begin{abstract}
Computational analyses of, e.g., genomic, proteomic, or metabolomic data, commonly result in one or more sets of candidate genes, proteins, or enzymes. These sets are often the outcome of clustering algorithms. Subsequently, it has to be tested if, e.g., the candidate gene-products are members of known metabolic processes. With OrfMapper we provide a powerful but easy-to-use, web-based database application, that supports such analyses. All services provided by OrfMapper are freely available at \url{http://www.orfmapper.com}.
\end{abstract}

\twocolumn

\section*{Introduction}

The amount of sequence related data increased dramatically during the past years.
This is due to improvements of high-throughput and computational methods in \textasteriskcentered omics that often yield long lists of gene, protein, or enzyme identifiers (IDs). In our laboratory we process different kinds of sequence based data, e.g., DNA-microarray derived \linebreak gene-expression data. The ultimate purpose of any gene-expression experiment is to produce biological knowledge. Independent of the methods used, the result of microarray experiments is, in most cases, a set of genes found to be differentially expressed between two or more conditions under study. The challenge faced by the researcher is to translate this list of differentially regulated genes into better understanding of the biological phenomena that generate \linebreak such changes. A good first step in that direction is the translation of the sequence ID list into a functional profile. Biological pathways can provide key information about the organization of biological systems. Major publicly available biological pathway diagram resources, including the Kyoto Encyclopedia of Genes and Genomes (KEGG) \cite{Kanehisa2004}, GenMAPP \cite{Dahlquist2002} and BioCarta\footnote{\url{http://www.biocarta.com}}, can be used to allocate sequence data in pathway maps. With this manuscript we do not intend to present a review about existing solutions but focus on our approach.

Our project requires the analysis of sequence cluster lists and extend the analysis to a maximum possible number of organisms. KEGG currently provides adapted maps for over 380 species covering the following molecular interaction and reaction networks: metabolism, genetic information processing, environmental information processing, cellular processes, human diseases.

In order to use the KEGG pathway database to display and map genes to KEGG pathways, we developed a web-based tool called OrfMapper. OrfMapper is an easy-to-use but powerful application that supports data analysis by extracting annotations for given keywords and gene, protein, or enzyme IDs, allocating these IDs to metabolic pathways, and displaying them on pathway maps. Two color codes can be assigned to the IDs, which can, e.g., represent sequence properties, organism identifiers, or cluster memberships. These color codes are used in the query output. The query results are displayed in hypertext format as a web page, prepared for download as tab-delimited raw text, and visualized on colored, hyperlinked KEGG metabolic pathway maps that can be downloaded in PDF format. Together with a version optimized for personal digital assistants, OrfMapper provides unique functionality with respect to accessing and displaying \linebreak KEGG pathway data.

\section*{Implementation}

\subsection*{Technical Background}
OrfMapper has been entirely developed with PHP version 4.3.4\footnote{\url{http://www.php.net}}, an open source scripting language that is especially \linebreak suited for Internet development. Creation of PDF is performed with FPDF version 1.53 \footnote{\url{http://www.fpdf.org}}, a freely available PHP class that allows generating PDF files. OrfMapper runs on a Apple Mac OS X version 10.2 operating system with an Apache version 1.3.33 HTTP server\footnote{\url{http://www.apache.org}}. The processed KEGG data are stored in a local relational MySQL database version 4.1.13 \footnote{\url{http://www.mysql.com}} database.

\subsection*{Database \& Updates}
The database behind OrfMapper contains gene identifiers, the annotation, organism, and pathway information, respectively. The database is updated monthly. Therefore, information from the KEGG FTP-server\footnote{\url{ftp://ftp.genome.jp/pub/kegg/}} and from the KEGG web site\footnote{\url{http://www.genome.ad.jp/kegg/}} are parsed.
In order to keep OrfMapper working and to avoid user query errors during updates, duplicated tables are used. Upon successful download and processing, the updated tables are activated  while outdated tables are inactivated.

\section*{Usage}

\subsection*{User Input}
OrfMapper was designed for prompt display of metabolic relations between gene products by the use of KEGG pathway maps.
A detailed online help guides the beginner through the user interface. The user has to specify either annotation keywords (e.g., "hydrogenase protein" or CoxA), gene IDs (e.g., KEGG, NCBI, UniProt), or enzyme IDs (i.e., EC-numbers). The user input can either be uploaded as an ASCII text file, be exported from spreadsheet applications (e.g., Microsoft Excel or \linebreak OpenOffice Calc), or directly pasted into a text area on the web page.

\subsubsection*{Data Format}
OrfMapper is made as flexible as possible in order to handle individual input data formats. The IDs can be listed either vertically or horizontally or mixed. They can be separated by all typical text delimiters, e.g., tabulators, spaces, commas and semicolons. Placing keywords in quotation marks forces \linebreak OrfMapper to perform a boolean AND query.

\subsubsection*{Organism Selection}
By default, all organisms are queried for all entered IDs and keywords. In order to restrict output to selected organisms, it is possible to specify those organisms in the first input row. This line must be preceded by an angle \linebreak bracket character "$>>$" followed by organism names or just parts of organism names (e.g., "droso" instead "Drosophila melanogaster"). The organism names must be separated by commas. If no match to an organism name is found, all organisms are queried.

\subsubsection*{Coloration}
In order to customize visualization, the user may specify colors for individual IDs. Therefore, either a color name (e.g., yellow, blue, red) or a hexadecimal RGB code (e.g., \#FFFF00) can be appended to IDs and keywords with two underscore characters "\hspace{1pt}\textunderscore{}\hspace{2pt}\textunderscore{}\hspace{1pt}" (e.g. genename\hspace{1pt}\textunderscore{}\hspace{2pt}\textunderscore{}\hspace{1pt}blue, genename\hspace{1pt}\textunderscore{}\hspace{2pt}\textunderscore{}\hspace{1pt}\#000080, keyword1\hspace{1pt}\textunderscore{}\hspace{2pt}\textunderscore{}\hspace{1pt}red, "keyword1 keyword2"\hspace{1pt}\textunderscore{}\hspace{2pt}\textunderscore{}\hspace{1pt}green). This colors the enzyme box corresponding to the ID on a KEGG pathway map. Likewise, the user can add one additional value to change the box border color. This is achieved by adding another color preceded by an underscore character to the ID (e.g., genename\hspace{1pt}\textunderscore{}\hspace{2pt}\textunderscore{}\hspace{1pt}blue\hspace{1pt}\textunderscore{}\hspace{2pt}\textunderscore{}\hspace{1pt}red). Coloration is extremely helpful to specify and, in the output, to identify gene products with common properties, such as expression levels or cluster affiliation.

\subsubsection*{Spreadsheet Import}
Large sets of query data are often stored in spreadsheet applications, e.g., Microsoft Excel, OpenOffice Calc, or Microsoft Access. Thus, we took special care to simplify date import from these applications. If the data are organized in three columns (ID, box color, and box border color, respectively), then they can directly copy-pasted into OrfMapper. Upon clicking the \textit{Convert Tab} button, all tabulators are converted to underscores, as required.

\subsection*{Output}
\begin{figure}[htb]
\centerline{\includegraphics[width=5.5cm]{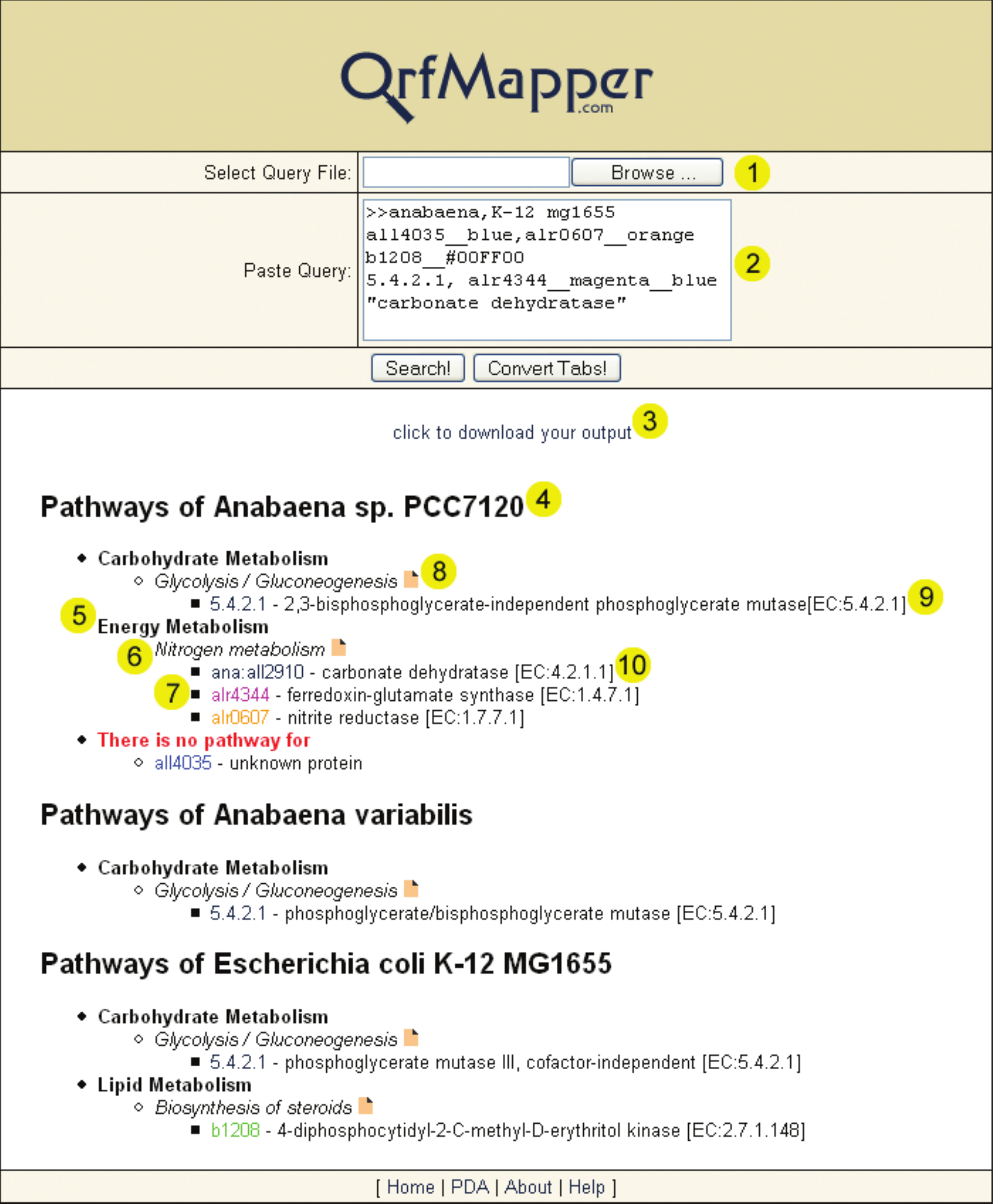}}
\caption{\textit{OrfMapper GUI. The query is either uploaded from a local file (1) or typed/pasted into the input field (2). Results are visualized as HTML or can be downloaded as tab delimited file (3). Hits are organized by organisms (4), metabolisms (5), submetabolisms (6), and enzymes (7). Pathway maps with colored hits can be downloaded as PDF (8) and gene information retrieved (9, 10).}}\label{fig1}
\end{figure}

OrfMapper creates three forms of output: hypertext, raw tab-delimited text, and graphical PDF pathway maps, respectively. The hypertext query result contains all gene annotations, pathway information, and hyperlinks to KEGG pathway maps corresponding to the user defined query (Fig. \ref{fig1}). This output is sorted by organism names, metabolic categories, pathways, and gene products. The latter two levels are hyperlinked to the corresponding KEGG information pages. This query result can be downloaded as raw tab-delimited text file for further processing. The first line of the text file contains the IDs given by the user. All following lines contain the full set of query results with the following entries: sequence or enzyme ID, KEGG species:sequence ID, annotation with EC-number and KEGG orthology ID, KEGG organism ID, species name, KEGG pathway map number, metabolic pathway name, box background color, and box border color. \linebreak Upon clicking the document symbol in the hypertext query results, OrfMapper creates a PDF version of the corresponding KEGG pathway map. The graphical PDF map can be saved locally, is scalable, optimized for printing, and includes hyperlinks to KEGG metabolite and enzyme information. If colors were assigned to sequence IDs in the query input, the background and borders of enzyme boxes are colored in the PDF maps. The PDFs are oriented such that the KEGG pathway maps fit perfectly either to portrait or landscape paper format.

\section*{Discussion}
OrfMapper was designed for displaying metabolic pathway oriented information of keywords and nucleotide, protein, or enzyme IDs of sequenced organism. Numerous visualization tools for analyzing biological data are available. OrfMapper fills a gap by providing quick access to pathway information via one input field with flexible input formats and output coloration options.

KEGG itself provides an integrated tool that can be used to color metabolic pathway objects . However, OrfMapper has a much broader functionality by allowing cross-species queries, giving a more detailed output, hyperlinking individual genes, and converting the colored pathway maps to PDF format retaining hyperlinks.

A condensed version of OrfMapper requiring less screen space and showing reduced output is devoted to palm-sized PDAs. Its screen size is scaled to 240 pixel width and the output of gene annotations is omitted. If equipped with WLAN, this allows on the spot information retrieval and mapping of keywords and gene or enzyme IDs, e.g., during research seminars.

OrfMappers' functionality will continuously be expanded. While the simple graphical user interface and query syntax will stay unchanged, extensions with respect to the application of functional characters are planned. We are currently integrating further sequence IDs, e.g., from the protein data bank (PDB). Furthermore, we are planning to facilitate nucleotide and protein sequence querying.

\section*{Acknowledgement}
This work is part of the BMBF funded Cologne University Bioinformatics Center (CUBIC). We like to thank Professor D.\ Tautz for generous support, Toshiaki Katayama from KEGG for \linebreak prompt help, and all beta testers for their valuable comments.
Conflict of interest: non declared


\end{document}